%Ref: MS 5203  Eric V. Linder 
%\input dspace.tex 
\magnification=1200
\def\pal{\parallel} 
\def\bi{\vec}
\def\dl{\delta}
\def\sg{\sigma}

\def\th{\theta}

\def\bP{\bi\Phi}
\def\bp{\bi\phi}
\def\bt{\bi\theta}
\def\bx{\bi\psi}
\def\tg{\sigma_\th}
\def\la{\langle}
\def\ra{\rangle}
\def\st{\sigma_t} 
\def\stt{\st^2} 
\def\sep{|\bP-\bx|^2} 
\def\vps{\bx} 
\def\vph{\bP}
\def\Th{\Theta} 
\def\bc{\vec D}

\leftline{\bf Correlated Gravitational Lensing of the Cosmic Microwave 
Background} 
\bigskip
\leftline{Eric V. Linder} 
\leftline{Imperial College, Prince Consort Road, London SW7 2BZ, 
England; el@ic.ac.uk} 
\bigskip
\bigskip 
\noindent{\bf Abstract.} 
Cosmological inhomogeneities gravitationally deflect radiation propagating 
from distant 
sources, transforming the spatial and angular correlation functions of 
intrinsic source properties.  For a gaussian 
distribution of deflections (e.g. from a primordial gaussian density 
perturbation spectrum or from the central limit theorem) 
we calculate the probability distributions for geodesic deviations.  If 
the intrinsic variable is also gaussian, e.g. the large scale velocity 
flow field or cosmic microwave background temperature anisotropies, then 
distributions and correlation functions of the observed image sky 
properties can be obtained.  Specialising to CMB temperature fluctuations 
we rederive simply the influence of independent gravitational lensing on 
the anisotropy angular correlation function and calculate the new effect 
of lensing correlated with the anisotropies, e.g. arising from the same 
primordial gravitational perturbation field.  Characteristic magnitudes 
and scales are given in terms of the density power spectrum.  The 
correlated deflection-temperature effect is shown to be negligible. 
\medskip 
\leftline{{\bf Key words:} Gravitational lensing, Cosmic microwave 
background, Cosmology: theory} 
\bigskip\medskip 

\leftline{\bf 1. Introduction} 
\medskip\noindent 
The origin and development of structure in the universe needs to be 
probed by means of observations of distant sources.  Not only the 
direct properties of those sources, however, but also the 
characteristics induced in the radiation propagating through the 
gravitational inhomogeneities of the structure provide clues to the 
nature and formation of structure.  Within linear theory, i.e. when 
the deviations from homogeneity are still perturbations, there are 
simple and intimate relations between the important cosmological 
descriptors of the matter, such as the large scale velocity flow field 
or the energy density fluctuations, and of radiation propagation 
characteristics such as temperature anisotropies (redshifts) and 
gravitational lensing deflections (momentum changes). 
These interrelationships constrain theories of structure 
formation. 

This paper 
concentrates on the statistical effects of the inhomogeneities on 
observations, developing an analytic approach that requires only that 
the physical processes behind the observable and lensing quantities 
be gaussian in nature.  This avoids sensitivity of results to a 
specific geometry or density perturbation model and is both physically 
relevant and
surprisingly powerful.  Although here we are primarily concerned with 
temperature anisotropies in the cosmic microwave background radiation 
(CMB), the method is easily generalised to investigations of the 
velocity flow field or density perturbations.  

Before primordial CMB anisotropies were first detected by the Cosmic 
Background Explorer satellite (COBE; Smoot et al. 1992) there was much 
discussion in the literature concerning the effects of inhomogeneities 
and their gravitational lensing on the anisotropies (Blandford \& 
Narayan 1992 provide a review with references).  More recently Fukushige, 
Makino, and Ebisuzaki (1994) investigated this by calculating 
scattering with an N-body code, and Tomita (1996) considered the 
effects of superhorizon scale inhomogeneities.  For the 
statistical distribution of the lensing deflections, in particular 
the correlation of neighboring lines of sight, crucial to the key 
property of geodesic deviation, various assumption have been made, 
ranging from a diffusion approximation 
to nearest neighbor correlations.  

No previous work, however, dealt with the case where the deflections 
are not superimposed independently on the temperature fluctuations 
but are physically correlated with them.  For example, consider a 
hot spot in the CMB.  This corresponds to a region where the primordial 
gravitational potential deviates strongly from the mean but that potential 
also determines the strength of the lensing caused by the primordial 
density field from that region.  Thus one can imagine that extremes of 
the temperature field are preferentially strongly lensed relative to 
the milder deviations, leading to a significant distortion of the intrinsic 
temperature correlation function.  

This paper 
investigates the properties of cosmologically important gaussian fields 
including the possibility of such a cross correlation.  The joint 
probability formalism allows for concise derivations of familiar 
results, e.g.~beam smearing, from gravitational lensing while showing 
new effects from the cross correlation.  By estimating the magnitude 
of these on anisotropy observations we can probe 
large scale structure in the universe by placing constraints on the 
density power spectrum $P_k$.  The conditional probability formalism 
also points out ways to test the 
underlying gaussian nature of the primordial cosmological perturbations. 

In \S2 we review the mathematical basis and derive 
properties of joint and conditional probability 
distributions for the variables entering the problem -- the relative 
deflections of the null geodesics and the intrinsic source sky 
characteristics.  These are combined in \S3 to form expressions for the 
observable correlation functions, especially under the condition of 
coherence between the source and propagation conditions.  
Section 4 relates the mathematical results to the underlying physics in 
terms of the density fluctuation power spectrum and \S5 presents 
quantitatively the effects on CMB anisotropy measurements. 
\bigskip 

\leftline{\bf 2. Probability distributions}
\medskip\noindent 
In an imperfectly homogeneous universe radiation observed at some angular 
position $\bp$ on the sky may have been deflected during propagation to us 
from a source whose 
true position is $\bx$.  By true position we mean that position at which the 
source would appear if the inhomogeneity (gravitational lens) were  
smoothed out.  A general line of sight has 
$$\bp=\vec\Psi+\bt\eqno(1)$$
for a sky projected deflection angle $\bt$, with $\bt$ varying along 
different directions.  The central question for observations in a 
universe with inhomogeneities is how to relate an intrinsic function 
of the source position to the observed function of image position. 

In order to investigate this in as model independent a way as possible, 
we assume that the gravitational lensing deflection and the physical 
source variables of interest individually have $d$ dimensional 
gaussian distributions with zero mean (no preferred direction).  Then 
$$p(\vec v)=[\pi(2\sg^2/d)]^{-d/2} e^{-v^2/(2\sg^2/d)},\eqno(2)$$ 
where the variance $\sg^2=\la v^2\ra$.  One 
justification for adopting a gaussian is that if the variable, e.g. 
projected deflection $\bt$, is compounded out of many elements, e.g. 
isolated spatial deflections at different distances from the observer 
(many inhomogeneities along the line of 
sight), then the central limit theorem leads to a gaussian distribution.  
Another rationale enters if the underlying physical process 
has a gaussian nature, generated by quantum fluctuations for example.  
One case would be gaussian density perturbations which create gaussian 
distributions in lensing deflection, velocity fluctuations, and 
temperature anisotropies in the linear regime.  This relation is discussed 
further in \S4. 

Since absolute shifts in the source position are undetectable we must 
compare at least two light rays, which requires knowing the joint 
probability of deflections along lines of sight 1,2,....  
More generally one can consider the variables $\vec v_i$ to denote 
different physical processes, such as density or temperature fields. 
Using the 
characteristic function approach (Cram\' er 1957) the joint probability 
of occurence is found to be 
$$p(\vec v_1,\vec v_2,...,\vec v_n)=(2\pi/d)^{-nd/2}M^{-d/2}\exp\bigl
\{-{d\over 2M}\sum_{i,j}M_{ij}\vec v_i\cdot\vec v_j\bigr\},\eqno(3)$$
where $M$ and $M_{ij}$ are the determinant and cofactors, respectively, of  
the covariance matrix composed of elements 
$m_{ij}=\la\vec v_i\cdot\vec v_j\ra$.  Angle brackets denote integration 
over the joint probability distribution.

In the case of two sky variables ($n=d=2$), for example comparing 
deflections of two rays, (3) simplifies to 
$$p(\bt_1,\bt_2)=\bigl[\pi^2\sg_1^2\sg_2^2(1-b^2)\bigr]^{-1} e^{-[(\th_1^2
/\sg_1^2)-2b(\bt_1\cdot
\bt_2/\sg_1\sg_2)+(\th_2^2/\sg_2^2)]/(1-b^2)},\eqno(4)$$
where the correlation coefficient $b=\la\bt_1\cdot\bt_2\ra/\sg_1\sg_2$.  
This distribution is normalized, and integration over one variable 
recovers the gaussian distribution for the other variable.  As $b\to0$ 
(incoherence or unrelated variables)  
the distribution separates into two independent gaussians while as $b\to1$ 
(total coherence) it approaches a delta function $\delta(\bt_2-\bt_1)$. 

One can imagine physical situations in which the variances $\sg_1$, 
$\sg_2$ of the two 
variables $\bt_1$, $\bt_2$ are not identical even though they correspond 
to the same physical process, e.g. gravitational deflection.  For example 
if the universe were not homogeneous then lines of sight through different 
patches of the universe might have different statistical properties.  
Even with homogeneity, consider sources at different redshifts. 
Because of the differing path lengths the cumulative gravitational 
deflections and hence variances $\sg_1$, $\sg_2$ would be unequal, as 
is evident from the explicit physical expressions (34) and (36). 
If we concentrate on the source being the CMB, however, the variances 
of deflections along any lines of sight are equal: $\la\th_1^2\ra= 
\la\th_2^2\ra=\sg_\th^2$. 

For other physical situations and gaussian 
processes, such as the large scale velocity field, evaluation out to 
different distances can be of interest, e.g. windowing different 
volumes of the density field or slicing redshift surveys.  Then, for 
example, one can find probability distribution functions such as for the 
ratio of the variable value in sample 1 to that in sample 2, 
$k=(\th_1/\sg_1)/(\th_2/\sg_2)$: 
$$p(k)=2(1-b^2)\,k^{-1}(k+k^{-1})\,[(k+k^{-1})^2-4b^2]^{-3/2}.\eqno(5)$$ 
As expected, the differential probability $p(k)\,dk$ is invariant 
under transforming $k$ to its reciprocal.  One can also calculate angular 
distributions, e.g. $p(\mu)$, where $\mu=\bt_1\cdot\bt_2/\th_1\th_2$. 
But this is of more use when dealing with three dimensional vector 
variables such as ${\vec v_R}$, the velocity flow field out to some survey 
depth $R$. 

Of particular physical interest is the difference between a quantity 
evaluated along two different lines of sight, e.g. the relative 
deflection or deviation $\bc=\bt_2-\bt_1$. 
Similarly one can investigate the coherence of the variables, e.g. 
$\beta=\bt_1\cdot\bt_2$.  Both are important to the physics. 
From (1) the observed separation of two rays $\bP=\bp_2-\bp_1$ is 
$$\bP=\bx+\bc,\eqno(6)$$
where $\bx=\vec\Psi_2-\vec\Psi_1$ is the intrinsic (true) separation and 
the separations can represent either the angular distance between 
two sources or the distance within a source (e.g. source size). 
Thus $\vec D$ is involved in the transformation of field separations 
and the quantities that depend on them. 
The coherence variable $\beta$ measures how independent the 
deflections are.  Physically, we expect that neighboring rays have strongly 
correlated deflections while lines of sight far apart have independent 
deflections.

To find the statistical distribution of the deviation and coherence one 
changes variables in (4) to, say, $\bc$, $\beta$, and $h=\th_2^2-
\th_1^2$, employing the Jacobian of the transformation: 
$$\eqalign{d^4P(\bc,\beta,h)&={\cal J}^{-1}\left({\bc,\beta,h\over \bt_1,
\bt_2}\right)\,p(\bt_1,\bt_2)\cr 
&=(\pi\tg)^{-2}Q^{-1} e^{-D^2/Q} e^{-2(1-b)
\beta/Q}(D^4+4\beta D^2-h^2)^{-1/2}\,\,d^2\bc\,d\beta\,dh,\cr}\eqno(7)$$
with $Q=\tg^2(1-b^2)$.  From this new joint distribution one derives 
the individual distributions by integrating over the other variables: 
$$\eqalignno{p(\bc)&=\bigl[2\pi\tg^2(1-b)\bigr]^{-1} e^{-D^2/2\tg^2
(1-b)},&(8)\cr 
p(\beta)&=\cases{\tg^{-2} e^{-2\beta/\tg^2(1+b)},&$\beta\ge0$,\cr 
\tg^{-2}e^{2\beta/\tg^2(1-b)},&$\beta\le0$,\cr}&(9)\cr 
p(h)&=\bigl[2\tg^2\sqrt{1-b^2}\bigr]^{-1} e^{-\vert h\vert/\tg^2
\sqrt{1-b^2}}.&(10)\cr}$$ 
Equations (8-10) illustrate that the distribution of a linear combination 
of gaussian variables 
is itself gaussian (Cram\' er 1957); note 
only $\bc$ is such a linear function.  The magnitude, or modulus, of a 
gaussian distributed vector quantity is Rayleigh distributed, 
$$p(D)=2\la D^2\ra^{-1}D\,e^{-D^2/\la D^2\ra}.\eqno(11)$$ 
We can find the moments of our chosen variables: 
$$\eqalignno{\la\bc\ra&=0=\la\bc^{2n+1}\ra\quad;\quad 
\la D^2\ra=2\tg^2(1-b)\quad;\quad\la D^{2n}\ra=n!\,
\la D^2\ra^n,&(12)\cr 
\la\beta\ra&=b\tg^2\quad;\quad\la\beta^n\ra=2^{-(n+1)}n!\,
\tg^{2n}\,\bigl[(1+b)^{n+1}+(-)^n(1-b)^{n+1}\bigr],&(13)\cr}$$ 
and verify that they have the expected behavior in the limits $b=0$ 
and $b=1$.  The above use of relative and coherence variables, Jacobians, 
and moments is equally applicable to variables besides the ray deflection. 

Note that since by (6) the observed separation on the sky depends 
only on the intrinsic separation and $\bc$, we can write correlations of 
position dependent quantities using only $p(\bc)$ and not the full 
$p(\bt_1,\bt_2)$.  For example for any function $f$ only of the relative 
field positions, e.g. separation (which is all that can enter in a 
globally homogeneous universe), 
$$\eqalign{\int d^2\bt_1\int d^2\bt_2\, f(\bt_2-\bt_1)\,p(\bt_1,\bt_2)&= 
\int d^2\vec A\int d^2\vec D\,f(\vec D)\,p(\vec D,\vec A)\cr &=\int 
d^2\vec D\,f(\vec D)\,p(\vec D),\cr}\eqno(14)$$ 
where, say, $\vec A=(\bt_1+\bt_2)/2$. 

We can immediately see that for the transformation of an intrinsic sky 
function $C_0$ of only the source separation $\psi$ into an observed 
function $C$ of the image separation $\Phi$, 
$$\eqalign{C(\Phi)&=\int d^2\bc\int d^2\bx\,C_0(\psi)\,p(\bc)\,\delta(\bP-
\bx-\bc)\cr &=\int d^2\bx\, C_0(\psi)\,p(\bc=\bP-\bx)\cr &=(\pi S^2)^{-1} 
\int d^2\bx\,C_0(\psi)\, e^{-(\bP-\bx)^2/S^2}\cr 
&=2S^{-2}\int_0^\infty d\psi\,\psi\, C_0(\psi)\,e^{-(\psi^2+\Phi^2)/S^2}\, 
I_0(2\psi\Phi/S^2),\cr}\eqno(15)$$
where $I_0$ is a modified Bessel function and $S^2=\la D^2\ra$.  

This recreates equation (24) of Wilson \& Silk (1981) for beam 
smearing, or scattering due to the observer and hence independent 
of source and path properties.  In this case $S$ is just a constant 
corresponding to the dispersion of the telescope beam response. 
The formalism epitomized by (15) also provides a compact 
method for deriving equation (19) of Linder (1990a) for the 
independent lensing of the microwave 
background temperature anisotropy correlation function.  Here the path 
properties do enter but are uncorrelated with the intrinsic source 
characteristics observed.  Then $S$ will be a function of path 
length and field separation $\Phi$, due to the different physical 
conditions along different lines of sight.  

The peak value of the 
correlation function, at $\Phi=0$, will be diluted by a factor 
$1+S^2/(2\Theta^2)$, where $\Theta$ is the characteristic scale, 
or coherence angle, of the observed quantity (cf.~equation 18).  Any 
oscillations in the angular behavior $C_0(\psi)$ will be damped by 
the coarse graining of the integral.  Observers seek to have beam 
sizes much smaller than the anisotropy scale of interest, $S\ll\Theta$, 
in order to avoid this smearing.  In the independent lensing case 
the zero lag correlation is unaffected since at $\Phi=0$ the lines 
of sight are identical, forcing $S=0$ ($b=1$).  Smearing does enter 
at larger angles though (cf.~Linder 1990a, Figures 2 and 3), its 
importance going as $S^2/\Theta^2\sim (\sigma_\theta/\Theta)^2(1-b)$. 

However, (15) only holds if the lensing process is independent of the 
observed process, i.e.~detected characteristic, so that the separation 
into $C_0(\psi)\,p(\vec D)$ is possible.  Section 3 provides more 
detailed discussion of this point and treats the 
correlated case, correcting (15) to (24), (25).  Meanwhile let us 
continue examining the necessary variables. 

The other gaussian variable to enter our correlated microwave background 
calculation is 
the fractional temperature anisotropy $t_i=\Delta T(\vec\phi_i)/T$.  This 
is a scalar variable ($d=1$) on the sky. 
Hence its probability function is 
$$p(t_1,t_2)=(2\pi\stt\sqrt{1-B^2})^{-1}e^{-(t_1^2-2Bt_1t_2+t_2^2)/
2\stt(1-B^2)},\eqno(16)$$ 
where the variance $\stt=\la t^2\ra$ and the correlation parameter 
$B=\la t_1t_2\ra/ \stt$.  In a more explicit notation, 
$$B\stt=\left\la{\Delta T\over T}(\vec\phi_1)\,{\Delta T\over T}(\vec 
\phi_2)\right\ra_{\Phi=|\vec\phi_1-\vec\phi_2|}=C(\Phi),\eqno(17)$$ 
the familiar angular anisotropy correlation function.  In this notation, 
$$\eqalign{\stt=&\la t^2\ra=C(0)\cr B&=\la t_1t_2\ra/\stt=C(\Phi)/C(0)\cr 
\Th&\equiv[-C''(0)/C(0)]^{-1/2}=(-d^2B(0)/d\Phi^2)^{-1/2},\cr}\eqno(18)$$ 
relating the correlation function $C$ and the coherence angle or curvature 
$\Th$ to the probability function characteristics $\st$ and $B$. 
Because the temperature is a random process evaluated at the same last 
scattering surface regardless of line of sight, the variance $\stt$ is 
not dependent on direction (cf. equation 31). 

While $B$ or $C(\Phi)$ gives the temperature anisotropy correlations 
averaged over the sky, i.e. the rms value, one can also calculate the 
entire probability distribution of the correlations from (16).  
For the coherence variable $c=t_1t_2$ this is 
$$p(c)=(\pi\stt\sqrt{1-B^2})^{-1}K_0[|c|/\stt(1-B^2)]\,e^{cB/\stt 
(1-B^2)},\eqno(19)$$ 
where $K_0$ is a modified Bessel function. 
This is important for computing the cosmic variance or intrinsic error 
due to the fact that we observe only a single universe, i.e. one 
realization of the probability distribution.  While the mean 
$\la c\ra=\stt B=C$, the variance is 
$$\sg_c^2=\la c^2\ra-\la c\ra^2=\st^4(1+B^2).\eqno(20)$$ 
Note that the distribution of $c$ is not gaussian. 

Observations of the CMB sky temperature can be related to $C$ according 
to the geometry 
of the fields, i.e. the number of telescope beams and the angles between 
them.  This can be made explicit by defining the differential variable 
$r=t_2-t_1$ with gaussian distribution 
$$p(r)=[4\pi\stt(1-B)]^{-1/2} e^{-r^2/4\stt(1-B)}.\eqno(21)$$ 
A two beam experiment with throw $\Phi$ measures the variance 
$$\la r^2\ra=2\stt(1-B_\Phi)=2\,[C(0)-C(\Phi)],\eqno(22)$$ 
while a three beam one 
symmetric about the center field measures the variance 
$$\eqalign{\la R^2\ra&=\la (r_a-r_b)^2\ra=2\la r^2\ra(1-b_{ab})\cr 
&=2\stt\,(3-4B_\Phi+B_{2\Phi})=6C(0)-8C(\Phi)+2C(2\Phi),\cr}\eqno(23)$$ 
where indices $a,b$ represent the two pairs of fields.  Note the 
similarity of form between (22) and (23): $\stt=\la t^2\ra\to\la r^2\ra$, 
$B\stt=\la t_1t_2\ra\to\la r_ar_b\ra=b_{ab}\la r^2\ra$. 

Rather than 
obtaining only the variance one could, if desired, compute other 
quantities such as the conditional probability of obtaining a certain value 
in one set of fields given that a fixed value was observed in other fields. 
This could be useful in that discrepancies between observations and 
probability theory could be a possible probe of 
nongaussianness in the temperature fluctuation distribution since only 
gaussian processes are fully determined by the variance. 

In summary, for the physical situation of gravitational inhomogeneity 
deflection of microwave background radiation we are interested in 
the mathematical correlations between two two dimensional 
gaussian variables -- one (the sky deflection $\bt$) a vector and one 
(the temperature fluctuations $t$) a scalar.  We will see that it is in 
fact convenient to decompose the vector deflection into arbitrarily 
oriented but orthogonal components, thus giving a total of three variables 
per field, all magnitudes. 
\bigskip 

\def\st{\sg}
\leftline{\bf 3. Cross correlations} 
\medskip\noindent 
The expression for the observed CMB temperature anisotropy correlation 
function will involve both the intrinsic temperature correlations and 
the conditional probability distribution of deflections given 
that the intrinsic fields have certain temperature fluctuations, i.e. 
values of the gravitational potential field. 

Mathematically, the intrinsic correlation function is defined by the 
integral of the field temperatures weighted by the joint probability of 
obtaining those particular values in each field, 
$$C_0(\psi)=\int dt_1\int dt_2\,p(t_1,t_2)\,t_1t_2,\eqno(24)$$ 
where $\vps$ is the field separation. 
To include the gravitational deflections one simply maps the intrinsic 
sky coordinates onto the observed ones because quasistatic gravitational 
lensing induces no redshift, i.e. the photon energy and hence temperature 
is unaffected.  The geometry of this mapping is illustrated in Figure 1. 

Now to calculate the observed correlation function one 
employs Bayes' theorem and simply adds up 
all the ways in which the fixed observed field separation $\vec\Phi$ can be 
generated from the various intrinsic field separations $\vec\psi$.  That is, 
one sums over all possible quadrilaterals with one fixed side, weighted by 
the probability for achieving that shape, i.e. obtaining deflections of 
the necessary magnitudes and directions, given the physical conditions 
represented by $t_1,t_2$. 

So the observed correlation function becomes 
$$C(\Phi)=\int d^2\vec D\int d^2\vec\psi\int dt_1\int dt_2\, p(t_1,t_2)
\,t_1t_2\,p(\vec D| t_1, t_2)\,\dl^{(2)}(\vec\Phi-\vec\psi-\vec D), 
\eqno(25)$$ 
where 
$\vec D=\bt_2-\bt_1$ is the relative deflection mapping $\vec\psi\to 
\vec\Phi$, the delta function enforces closure of the quadrilateral, 
and $p(\vec D|t_1,t_2)$ is the conditional probability.  
This is a direct translation of the above words into an equation.  
For example the phrase ``given the physical conditions'' indicates that the 
probability function of all the variables does not range over the whole 
parameter space but is constrained by some prior knowledge or condition; 
hence one uses the conditional probability distribution to reflect this. 
Now conditional probability is calculated by the full probability divided 
by the probability to achieve the set condition.  Hence we can 
convert the product of the temperature probability function $p(t_1,t_2)$ 
and the conditional deflection probability function given those 
temperature values $p(\vec D|t_1,t_2)$ 
to a joint probability over all the variables $p(\vec D,t_1,t_2)$. 

However, the expression is still 
not that useful for comparison with observations as it requires an 
integration 
over the ensemble of probability realizations, yet we have only a single 
universe to work with.  The solution is to employ the ergodic hypothesis 
which states that we can substitute an integration over the whole sky 
for an ensemble average.  Thus we can replace integration over all the 
values which the random process could take on at a location by 
integration over all evaluation locations.  

The final form of the 
observed correlation function is then 
$$\eqalign{C(\Phi)&=\int d^2\vps\int dt_1\int dt_2\,t_1t_2\,p(\vec D=
\vph-\vps,t_1,t_2)\cr 
&=\int d^2\vec\psi\int d^2\vec\psi_1\,t(\vec\psi_1)\,t(\vec 
\psi_1+\vps)\,p(\vec D=\vph-\vps,t_1,t_2),\cr}\eqno(26)$$ 
enforcing $\vec\psi_2=\vec\psi-\vec\psi_1$ in the second line.  Only in 
the case where the 
deflection process is independent of the temperature fluctuations does 
the joint probability separate into a product of $p(\vec D)\,p(t_1,t_2)$, 
necessary to obtain the explicit appearance of the intrinsic correlation 
function $C_0(\psi)$ in (15). 

To obtain the joint probability function between the gaussian variables 
of the ray deflections and the sky temperatures, it is convenient to 
decompose $\vec D$ into orthogonal 
components.  This both puts all variables on an equal dimensional footing 
and makes the correlation $\la D_xD_y\ra=0$. 
By using the relative variable 
$\vec D=\bt_2-\bt_1$ we no longer explicitly need the deflection 
correlation parameter $\beta=\bt_1\cdot\bt_2$.  In \S4 we find that only 
the component of $\vec D$ along the field separation has a nonvanishing 
correlation with the field temperatures, so we choose our decomposition 
parallel and perpendicular to this axis. 

Now the joint probability function for the variables $D_\pal,D_\perp, 
t_1,t_2$ follows from (3) via the covariance matrix, 
$${\bf M}(D_\pal,D_\perp,t_1,t_2)=\pmatrix{s^2&0&-bs\st&bs\st\cr 
0&s^2&0&0\cr -bs\st&0&\sg^2&B\st^2\cr 
bs\st&0&B\st^2&\sg^2\cr}.\eqno(27)$$ 
One can read off from the matrix that 
$s^2$ is the variance of each component of the deflection $\vec D$, 
$\stt$ is now the variance of the temperature fluctuations, $B$ is the 
temperature correlation parameter, and $b$ is now the cross correlation 
parameter between the gravitational deflection and the temperature 
fluctuation. 

The joint probability distribution is then 
$$\eqalignno{p(D_\pal,D_\perp,t_1,t_2)&=(2\pi s\sg)^{-2}(1+B)^{-1/2}
(1-B-2b^2)^{-1/2}\,\times\cr 
&\qquad\qquad\exp \{-E/[2s^2\sg^2(1+B)(1-B-2b^2)]\},&(28)\cr 
E&=s^2(1-b^2)\,(t_1^2+t_2^2)-2s^2(B+b^2)\,t_1t_2-2bs\sg(1+B)
\,D_\pal(t_2-t_1)\cr 
&\quad +\sg^2(1-B^2)\,D_\pal^2+\sg^2(1+B)(1-B-2b^2)\,D_\perp^2.\cr}$$ 
Carrying out the integrations specified in (26) yields the observed 
CMB temperature anisotropy correlation function 
$$\eqalignno{C(\Phi)&=(2\pi s^2)^{-1}\int d^2\vps\,C_0(\psi)\,e^{-\sep/ 
2s^2}\cr &\qquad +(2\pi s^2)^{-1}\int d^2\vps\,b^2\,C_0(0)\,\Bigl[1-
\{(\vec\Phi-\vec\psi)\cdot\hat\psi\}^2/s^2\Bigr]\, e^{-\sep/2s^2}\cr 
&\equiv C_{ind}+C_{cor}.&(29)\cr}$$ 

The first term is proportional to $B=C_0(\psi)/\stt$ and is the 
contribution to the observed temperature 
correlations when the gravitational lensing is independent of 
the temperature anisotropies, i.e. there is no cross correlation.  This 
is precisely the old result derived in Linder (1990a) and mentioned in 
(15).  The second term, involving $b$, 
represents the new effect of including correlations 
between the gravitational deflection field and the temperature fluctuation 
field, for example due to both originating 
from the same primordial gravitational potential perturbation. 

The two crucial ingredients are thus the correlations $B$ [or $C_0(\psi)$] 
and $b$.  The former is a scalar function of field separation, 
i.e. independent of orientation, and from its definition one expects the 
latter to vary with angle as $\hat\Phi\cdot\hat\psi$; both are verified 
in the next section.  Then the angular part of the 
integrals can be carried out without knowing their explicit form: 
$$\eqalign{C(\Phi)=&s^{-2}e^{-\Phi^2/2s^2}\int_0^\infty d\psi\,\psi\, 
[C_0(\psi)\,I_0(\Phi\psi/s^2)+C_0(0)\,\beta^2(\psi,\Phi)\,Y]\,e^{-
\psi^2/2s^2}\cr 
Y&=[(\Phi^{-1}+2\Phi s^{-2})\,\psi+(2\Phi+4s^2\Phi^{-1})\,\psi^{-1} 
+6s^4\Phi\,\psi^{-3}]\,I_1(\Phi\psi/s^2)\cr 
&\quad -[s^{-2}\psi^2+(1+\Phi^2 s^{-2})+3s^2\psi^{-2}]\,I_0(\Phi\psi/s^2), 
\cr}\eqno(30)$$ 
where $I_\nu$ are modified Bessel functions and $\beta=b/(\hat\Phi\cdot
\hat\psi)$. 

The prescriptions for the variable variances and correlation parameters are 
given by the physics underlying the random processes and can be written 
in terms of the gravitational potential perturbations, or more 
conventionally the primordial density fluctuation power spectrum, as in 
the next section. 
\bigskip 

\leftline{\bf 4. Density fluctuation spectrum}
\medskip\noindent 
The mathematical basis of the statistical moment and correlation analysis 
relied on 
the gaussian nature of the observational characteristics.  This implied 
that all the physics resided in the variance and cross correlation of 
the characteristic variables.  Now we investigate those physical processes 
to obtain expressions for these quantities in terms of the underlying 
mechanism -- fluctuations in the cosmological density field -- expressed in 
terms of the density power spectrum. 

First consider the source property of the cosmic microwave radiation 
background temperature anisotropies.  The temperature fluctuations, or 
shifts in the photon energies, are induced by variations in the 
gravitational potential at the last scattering surface, $t=\phi/3$, 
(variations along 
the photon path contribute negligibly due to their quasistatic nature). 
In turn, the potential can be related to the density perturbations via 
the Poisson equation, reading $\phi_k=(3/2)H_0^2\,k^{-2}\dl_k$ in Fourier 
space.  Thus the variance of the temperature fluctuations can be written 
in terms of the density power spectrum $P_k=|\dl_k|^2$ as 
$$\sg^2=\la t^2\ra=C_0(0)=(8\pi^2)^{-1} H_0^4\int_0^\infty dk\,k^{-2}P_k.
\eqno(31)$$ 

The correlation parameter is 
$$B=\la t_1t_2\ra/\la t^2\ra=\int_0^\infty dk\,k^{-2} {\sin kL\Phi\over kL
\Phi} P_k\Big/\int_0^\infty dk
\,k^{-2}P_k,\eqno(32)$$ 
where $L=2H_0^{-1}[1-(1+z)^{-1/2}]$ is the comoving distance to the last 
scattering surface at redshift $z$.  (The small angle approximation 
$\Phi\ll1$ is used throughout.)  From (18) the coherence angle is then 
$$\Theta^2_t=3L^{-2}\int_0^\infty dk\,k^{-2}P_k\Big/\int_0^\infty dk\,
P_k,\eqno(33)$$ 
or $\Theta_t\approx1/(kL)$. 

For the ray deflection variables, the physical basis is gravitational 
lensing due to the density fluctuations along the line of sight.  
Following Linder (1990b) the geodesic equation gives 
$$\vec\th=2L^{-1}\int_0^{L} dx\int_0^x dx'\,\vec\nabla_\perp\phi,\eqno(34)$$ 
i.e. the deflection is caused by gradients in the gravitational 
potential.  The equation of geodesic deviation governs the relative 
deflection, or deviation, of neighboring rays: 
$$\vec D=\vec\th_2-\vec\th_1=2L^{-1}\int_0^L dx\int_0^x dx'\,x'\,(\vec\Phi 
\cdot\vec\nabla_\perp)\,\vec\nabla_\perp\phi.\eqno(35)$$ 

Converting the gravitational potential to the density power spectrum 
as before, by $\phi=(2\pi)^{-3/2}\int d^3k\,\phi_k e^{i{\bf k\cdot x}} 
=(3/2)(2\pi)^{-3/2}H_0^2\int d^3k\,k^{-2}\delta_ke^{i{\bf k\cdot x}}$, 
one then carries out the derivatives and integrations in (34) and (35). 
To calculate a variance one does the angle and ensemble averages as 
represented by the angle brackets to obtain for the variance of 
a single ray deflection 
$$\sg_\th^2=\la\th^2\ra=(3/2\pi) H_0^4L\int_0^\infty dk\,k^{-1}P_k,
\eqno(36)$$ 
and for the variance of the ray bundle deviation 
$$S^2=\la D^2\ra=2s^2=2\sg_\th^2\,(1-b_\th),\eqno(37)$$ 
in agreement with (12).  The deflection correlation parameter comes 
from (35)-(37): 
$$\eqalign{b_\th \sg_\th^2&=\int_0^\infty dk\,k^{-1}P_k\int d^2 
\vec\Omega\,f(\vec\Omega,\vec\Omega',kL)\cr 
&=(3/2\pi) H_0^4L\int_0^\infty dk\,k^{-1}\bigl[1-
(kL\Phi)^2/40+{\cal O}(kL\Phi)^4\bigr]\,P_k,\cr}\eqno(38)$$ 
where the angular variables $\vec\Omega,\vec\Omega'$ of $\vec k$ 
relative to the two lines of sight are related by 
a rotation of the axes by angle $\Phi$.  For example the direction 
cosine of the wave vector with respect to the second line of sight is 
$\mu'=\mu\,\cos\Phi+\sqrt{1-\mu^2}\sin\Phi\,\cos\beta$ with $\mu$ the 
cosine with respect to the first line of sight and $\beta$ the azimuthal 
angle in the first frame.  

Unfortunately the general evaluation of (38) 
is not analytically amenable but the second line of the equation shows 
the behavior for angles $\Phi\ll(kL)^{-1}$.  This is sufficient to find 
the coherence angle and other interesting physical parameters like the 
gravitational lensing ray crossing variable ${\cal D}^2=\lim_{\Phi\to0} 
\la D^2\ra/\Phi^2$ $=(3/40\pi)H_0^4L^3\int dk\,k\,P_k$.  When ${\cal D}^ 
2>1$ then the Jacobian of the sky-image map goes singular, rays cross, 
and the image appears grossly distorted or multiple. 
The coherence angle of the lensing deflections is defined analogously 
to (33), 
$$\Th_\th^2=20L^{-2}\int_0^\infty dk\,k^{-1}P_k\Big/\int_0^\infty dk\,k 
P_k,\eqno(39)$$ 
or $\Th_\th\approx1/(kL)$. 

For the cross correlation parameter $b$ one takes $t=\phi/3$ and 
(35) for $\vec D$, both involving Fourier transforms of $\phi_k$ 
and hence $\delta_k$, and carries out the same ensemble average 
calculational process.  This gives 
$$bs\sg=\la D_\parallel t\ra=-(3/4\pi^2) H_0^4\Phi\alpha\int_0^\infty 
dk\,k^{-2} 
P_k \,[1-(1/2)k^2L^2\psi^2+{\cal O}(kL\psi)^4],\eqno(40)$$ 
with $\alpha=\hat\Phi\cdot\hat\psi$ and again the general behavior 
unamenable.  In the limit of small 
field separation $s\sim\Phi$ (see equations 37, 38) so 
$$b\approx -(120/\pi)^{1/2}\alpha L^{-3/2}\left[\int_0^\infty dk\,k^{-2} 
P_k\Big/\int_0^\infty dk\,kP_k\right]^{1/2}.\eqno(41)$$ 
For the larger 
angle behavior one can turn to the more physical intuitive (though not 
more calculationally tractable) expression 
$$b^2=\left[\la D^2t^2\ra-\la D^2\ra\la t^2\ra\right]/(2s^2\stt)=\la 
D^2t^2\ra/\la D^2\ra\la t^2\ra-1,\eqno(42)$$ 
which follows from the joint probability function.  This shows that 
indeed $b$ vanishes as the deflections and temperature 
fluctuations become uncorrelated. 

To obtain specific quantitative results for the observed microwave 
background temperature anisotropy correlation function requires becoming 
more model 
dependent, choosing a particular density power spectrum.  
We illustrate the orders of magnitude for the variables and the behavior 
of the resulting correlation function by adopting a cold dark matter 
spectrum normalized to the detected microwave background quadrupole 
anisotropy, $P_k=1.8\times10^{-9}(a_2/2.2\times 10^{-5})^2 H_0^{-4}$ 
$k/[1+l_1k+l_2k^{3/2}+l_3k^2]^2$, where $l_1,l_2,l_3$ are length constants 
(Davis et al. 1985).  The results are $\st=1.1\times10^{-5}$, $\Th_t=14'$, 
$\sg_\th=3'$, $\Th_\th=5'$, and $b=-2\times10^{-4}$.  This value for $b$ 
is at $\Phi=0$ but its maximum at finite $\Phi$ is only a few times 
larger.  In terms of approximating the 
coherence angles by $(kL)^{-1}$, the characteristic wavenumber is 
$k\approx(15\,h^{-1}Mpc)^{-1}$. 
\bigskip 

\leftline{\bf 5. Conclusion} 
\medskip\noindent 
We have investigated the behavior of the observed CMB temperature 
anisotropy correlation function 
in a universe with density inhomogeneities 
causing a gaussian light deflection field, both correlated with and 
independent of the perturbations giving rise to the temperature 
fluctuations.  The ratio of the correlated effect to the independent 
one is of order the cross correlation parameter squared, $b^2$.  While 
one might expect that the deflection and temperature be strongly 
coupled, $b\approx1$ (cf. equation 42), this is found not to be true 
here.  In the case of gravitational lensing of photons from the last 
scattering surface the correlation influence on CMB anisotropies is 
negligible because one loses the path length ``resonance'' of the 
deflections and temperature fluctuations together tracing the gravitational 
potential field along the entire line of sight.  Such a dependence 
would increase the correlation parameter by a factor of $\approx kL
\approx 10^{2.5}-10^{3.5}$, making the effect significant.  Due to 
the quasistatic nature of the potential along the path, however, this 
does not give rise to observable temperature fluctuations.  Exceptions 
occur only in localized events such as cluster formation (Rees \& Sciama 
1968) or hot cluster cores (Sunyaev \& Zel'dovich 1970), which do not 
offer the opportunity for path resonance or gaussian statistics.  

Although the correlation of the two density perturbation derived 
variables considered here can be neglected, the formalism of calculating 
probability distributions and covariances of gaussian fluctuations is 
powerful.  Insights garnered here into the methods and relations may 
prove useful in other cosmological applications such as analyzing large 
scale velocity flows. 
\bigskip 

\noindent{\it Acknowledgments.} 
I thank Jacob Sharpe for a careful reading of the paper and the referee, 
Jim Bartlett, for improvements in clarity. 
\medskip 
\leftline{\bf References}
\smallskip 
\parindent=0in
Blandford, R.D., Narayan, R., 1992, ARA\&A, 30, 311
 
Cram\'er, H., 1957, Mathematical Methods of Statistics, Princeton 
University Press, Princeton 

Davis, M., Efstathiou, G., Frenk, C.S., White, S.D.M., 1985, 
ApJ, 292, 371 

Fukushige, T., Makino, J., Ebisuzaki, T., 1994, ApJ, 436, L107 

Linder, E.V., 1990a, MNRAS, 243, 353 

Linder, E.V., 1990b, MNRAS, 243, 362 

Rees, M.J., Sciama, D.W., 1968, Nat, 217, 511

Smoot, G.F. et al., 1992, ApJ, 396, L1 

Sunyaev, R.A., Zel'dovich, Ya.B., 1970, Ap. Sp. Sci., 7, 3

Tomita, K., 1996, ApJ, 461, 507

Wilson, M.L., Silk, J., 1981, ApJ, 243, 14
\medskip 
\leftline{\bf Figure caption}
\smallskip\noindent 
{\bf Fig.~1.} The geometry of the intrinsic field separation vector 
$\vec\psi$, the deflection vectors $\vec\th$ for the two lines of 
sight, and the observed separation vector $\vec\Phi$ is shown in the 
plane of the sky. 
\vfill\break
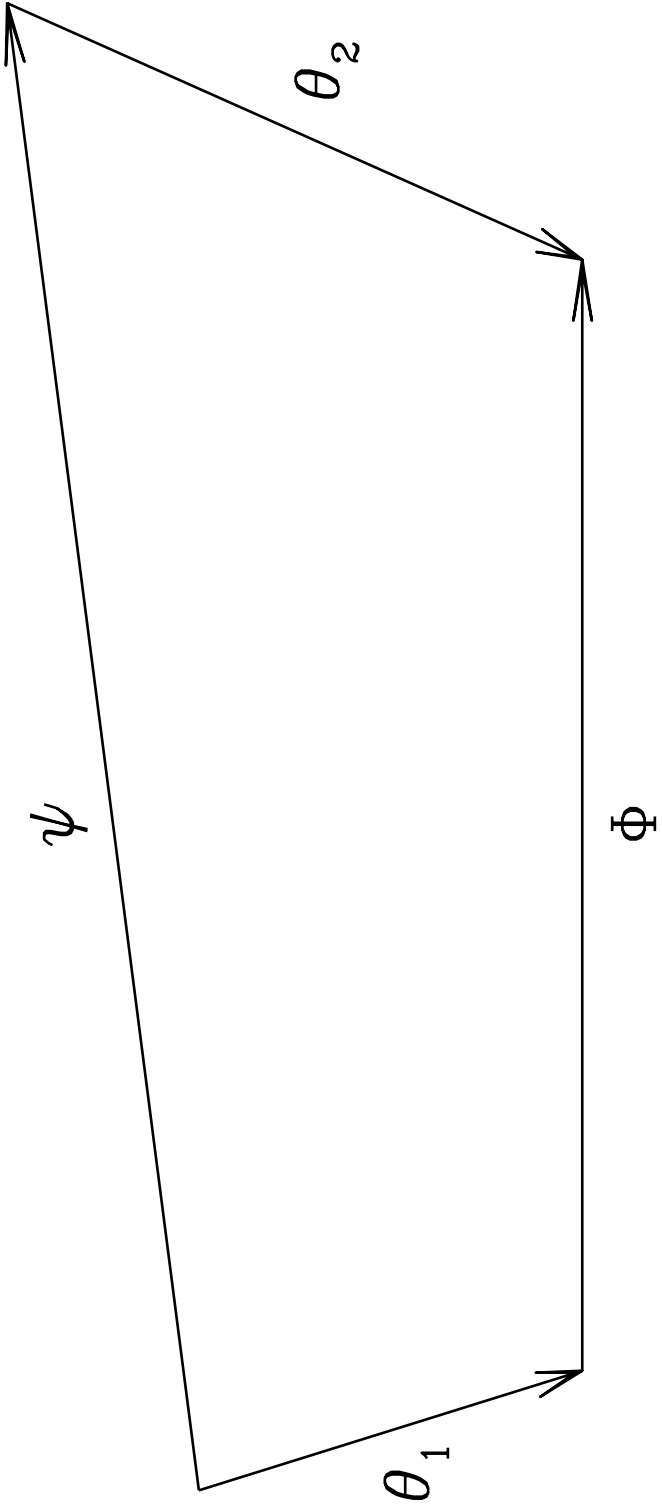
\bye